\documentclass[conference]{IEEEtran}
\usepackage {amssymb}
\usepackage {amsmath}
\usepackage {makeidx}
\usepackage [dvips]{graphicx,color}
\usepackage{epstopdf}
\usepackage{subfigure}
\usepackage{citesort}

\newtheorem{thm}{Theorem}

\newtheorem{ex}{Example}
\newtheorem{de}{Definition}

\newtheorem{lem}{Lemma}
\newtheorem{rem}{Remark}
\renewcommand{\QED}{\hfill $\Box$}

\newcommand{\pr}[1]{{\mathsf{Pr}}\left\{#1\right\}}

\renewcommand{\c}[1]{{\cal #1}}

\newcommand{\bb}[1]{{\mathbb #1}}

\newcommand{\DEF}{\stackrel{\rm def}{=}}
\renewcommand{\pr}[1]{{\sf Pr}\left\{ #1 \right\}}

\newcommand{\PScs}[1]{{\sf PS}_{C*}( #1 )}
\newcommand{\PSsm}[1]{{\sf PS}_{*M}( #1 )}
\newcommand{\PScm}[1]{{\sf PS}_{CM}( #1 )}
\newcommand{\IND}[1]{{\sf IND}( #1 )}

\renewcommand{\SS}[1]{{\sf SS}( #1 )}

\newcommand{\enc}{{\sf Enc}}

\newcommand{\dec}{{\sf Dec}}

\makeatletter
\def\eqnarray{%
        \stepcounter{equation}%
        \let\@currentlabel=\theequation
        \global\@eqnswtrue\global\@eqcnt\z@
        \tabskip\@centering
        \let\\=\@eqncr
        $$\halign to \displaywidth\bgroup\@eqnsel\hskip\@centering
        $\displaystyle\tabskip\z@{##}$&\global\@eqcnt\@ne
        \hfil$\displaystyle{{}##{}}$\hfil
        &\global\@eqcnt\tw@$\displaystyle\tabskip\z@{##}$\hfil
        \tabskip\@centering&\llap{##}\tabskip\z@\cr}
\makeatother

\ifCLASSINFOpdf
\else
\fi
\begin{document}
%
\title{Security Notions \\ for Information Theoretically Secure Encryptions}

\author{\IEEEauthorblockN{Mitsugu Iwamoto}
\IEEEauthorblockA{Center for Frontier Science, \\
the University of Electro-Communications\\
1--5--1 Chofugaoka, Chofu-shi, Tokyo, 182--8585 Japan\\
Email: mitsugu@inf.uec.ac.jp}
\and
\IEEEauthorblockN{Kazuo Ohta}
\IEEEauthorblockA{Graduate School of Informatics and Engineering, \\the University of Electro-Communications\\
1--5--1 Chofugaoka, Chofu-shi, Tokyo, 182--8585 Japan\\
Email: ota@inf.uec.ac.jp}
}


%


\maketitle

\begin{abstract}
This paper is concerned with  several security notions for information theoretically secure encryptions defined by the variational (statistical) distance. To ensure the perfect secrecy (PS), the mutual information is often used to evaluate the statistical independence between a message and a cryptogram. On the other hand, in order to recognize the information theoretically secure encryptions and computationally secure ones comprehensively, it is necessary to reconsider the notion of PS in terms of the variational distance. However, based on the variational distance, three kinds of definitions for PS are naturally introduced, but their relations are not known. In this paper, we clarify that one of three definitions for PS with the variational distance, which is a straightforward extension of Shannon's perfect secrecy, is stronger than the others, and the weaker two definitions of PS are essentially equivalent to the statistical versions of indistinguishability and semantic security. 
\end{abstract}


%
\IEEEpeerreviewmaketitle

\section{Introduction}

{\em Perfect secrecy} (PS)
is a strong security notion which is secure against an adversary 
with unbounded computing power. 
Perfect secrecy was defined by Shannon \cite{S-milestone}, and he proved that perfect secrecy is achieved by one time pad (Vernam) cipher \cite{Vernam-jaiee26}. 
Furthermore, in order to achieve perfect secrecy, Shannon also proved in \cite{S-milestone} that the entropy of a key must be greater than the entropy of a message, which makes perfect secrecy quite impractical. 

Roughly speaking, PS is defined by the statistical independence between a message $M$ and a cryptogram $C$. Specifically, we often require {\em almost} statistical independence between $C$ and $M$ to ensure PS. We note here that two metrics can be used to measure the almost statistical independence, i.e., the mutual information and the variational (statistical) distance. In general, the mutual information is often used in information theoretic cryptography since it guarantees stronger security compared to the security notions based on the variational distance due to Pinsker's inequality. 
On the other hand, the variational distance is often used in computationally secure cryptography: For instance, indistinguishability (IND) and semantic security (SS) are defined in terms of the variational distance. We note that several researchers recently discussed one time pad cipher under the security notions developed in computationally secure cryptography. For instance, Russell--Wang \cite{RW-it06} introduced {\em entropic security} based on semantic security, and they succeeded in shortening the key length of a symmetric key cryptosystem which is secure against an {\em unbounded} adversary. In addition, Dodis--Smith \cite{DS-tcc05} introduced another security notion which is closely related to indistinguishability, and they gave the other realization of entropic security by using extractors \cite{K-crypto94}. 

Given the above backgrounds, we are interested in PS defined by the variational distance, and its relation to IND and SS, which will be some help for comprehensive understanding of information theoretically secure encryptions and computationally secure ones. However, as we will see in Definition \ref{de:PSve}, three kinds of definitions of PS denoted by $\PSsm{\varepsilon}$, $\PScs{\varepsilon}$, and $\PScm{\varepsilon}$ can be naturally introduced in terms of the variational distance. It is obvious that these three notions of PS are the same when $\varepsilon =0$. However, in the case of $\varepsilon > 0$, their relations are not known. In this paper, we will point out that $\PSsm{\varepsilon}$ is stronger than the others by showing a pathological example. 
Furthermore, it will be proved that the remaining two definitions $\PScs{\varepsilon}$ and $\PScm{\varepsilon}$ guarantee essentially the same security as the statistical versions of IND and SS. 

The rest of this paper is organized as follows: In Section \ref{sec:preliminary}, notations and three variations of PS are introduced. Statistical IND is introduced in section \ref{sec:ind}, and the relations between PS and statistical IND are clarified. A relation between statistical IND  and statistical SS is proven in Section \ref{sec:ss}. Finally, a gap between one of three variations of PS and the other security notions are pointed out in Section \ref{sec:gap}. Technical lemmas are provided in Appendix. 

\section{Preliminaries}\label{sec:preliminary}

Let $M$, $K$, and $C$ be random variables taking values in finite sets $\c{M}$, $\c{K}$, and $\c{C}$, which correspond to sets of messages, keys, and cryptograms, respectively. For a random variable $X$ taking values in a finite set $\c{X}$ and an element $x \in \c{X}$, denote by $P_X(x)$ a probability of $X=x$. Let $\c{P}(\c{X})$ be the totality of probability distributions over $\c{X}$.

A symmetric key cryptography $\Sigma$ consists of a probability distribution $P_K\in\c{P}(\c{K})$ of a key, and a pair of an encryption function $\enc : \c{M} \times \c{K} \rightarrow \c{C}$, and a decryption function $\dec : \c{C} \times \c{K} \rightarrow \c{M}$, i.e., $\Sigma\DEF(P_K,\enc,\dec)$. Note that $K$ is chosen  independently of a message $M$, and $\enc$ and $\dec$ are deterministic maps. Suppose that a message is generated according to a probability distribution $P_M \in \c{P}(\c{M})$. Then, the probability distribution $P_C$ of a cryptogram is determined by $P_M$, $P_K$ and $\enc$. 
Let $P_{CM}$ be a joint probability distribution of a cryptogram $C$ and a message $M$, and denote by  $P_{C|M}$ a conditional distribution of a cryptogram when a message is given. 
Denote by $\mathbb{P}_{C|M}$ an $|\c{C}| \times |\c{M}|$ transition probability matrix\footnote{$|\cdot|$ denotes the cardinality of a set.} associated with $\{P_{C|M}(c|m)\}_{c\in\c{C},m\in\c{M}}$, i.e., each element of $\mathbb{P}_{C|M}$ corresponds to  $P_{C|M}(c|m)$ for $c \in \c{C}$ and $m \in \c{M}$. 
The following theorem states fundamental properties of $P_{C|M}$ for symmetric key encryptions. The proof is provided in Appendix \ref{app:fundamental-relation}.

\begin{thm}\label{thm:fundamental}
If a key $K$ is chosen independently of a message $M$, it holds that\footnotemark\footnotetext{$\pr{\cdot}$ is a probability with respect to a (joint) probability distribution of random variable(s) between the parentheses.}
\begin{eqnarray}\hspace*{-2mm}
 \forall c \in \c{C}, \forall m \in \c{M},~
P_{C|M}(c|m)=\pr{\enc(m,K)=c}.
\label{eq:fundamental-relation}
\end{eqnarray}

Furthermore, in the case of $|\c{C}|=|\c{M}|$, 
there exists a symmetric key cryptosystem $\Sigma$ satisfying \eqref{eq:fundamental-relation} 
iff (if and only if) the probability transition matrix $\bb{P}_{C|M}$ is {\em doubly stochastic}\footnotemark\footnotetext{A probability transition matrix $\bb{P}_{C|M}$ is {\em doubly stochastic} iff  $\sum_{c\in\c{C}}P_{C|M}(c|m)=\sum_{m\in\c{M}}P_{C|M}(c|m)=1$ holds.}. \\
 \QED
\end{thm}

Hence, we assume that the conditional probability distribution $P_{C|M}(c|m)$, $c\in\c{C}$, $m \in \c{M}$ is naturally defined by \eqref{eq:fundamental-relation} if a symmetric key cryptosystem $\Sigma$ is given. 

Shannon defined the notion of {\em perfect secrecy} as follows: 
\begin{de}[Perfect secrecy, \cite{S-milestone}]\label{de:PS}
A symmetric key cryptosystem $\Sigma= (P_K,\enc,\dec)$  guarantees 
{\em perfect secrecy} if 
\begin{eqnarray}\label{eq:PS}
\forall c \in \c{C},~ \forall m \in \c{M},~ P_{M|C}(m|c) = P_M(m) 
\end{eqnarray}
is satisfied for {\em arbitrary} message distribution $P_M$. \QED
\end{de}

Definition \ref{de:PS} means that no information of a message can be obtained from a cryptogram since {\em a priori} probability distribution $P_M$ of a message coincides with {\em a posteriori} probability distribution of $M$ computed by an adversary using a  cryptogram. 

It is easy to see that \eqref{eq:PS} is equivalent to 
\begin{eqnarray}\label{eq:PSm}
\forall c \in \c{C},~ \forall m \in \c{M},~P_{C|M}(c|m) &=& P_C(c)\\
\label{eq:PScm}
\forall c \in \c{C},~ \forall m \in \c{M},~P_{CM}(c,m) &=& P_C(c)P_M(m)
\end{eqnarray}
since \eqref{eq:PS} means that random variables $M$ and $C$ are statistically independent. 

We are now consider {\em relaxed} definitions of perfect secrecy.  
That is, we  define {\em almost} independence between a message $M$ and a cryptogram 
$C$ given by \eqref{eq:PS}--\eqref{eq:PScm} in terms of the variational (statistical) distance\footnotemark\footnotetext{For two probability distributions $P_X$, $P_Y$ over a finite set $\c{A}$, the variational distance $d(\cdot,\cdot)$ is defined by $
d(P_X,P_Y)\DEF 
(1/2)\sum_{a \in \c{A}}| P_X(a) - P_Y(a)|
=
\max_{f: \c{A} \rightarrow \{0,1\}}|\pr{f(X)=1} -  \pr{f(Y)=1} |.$
} denoted by $d(\cdot,\cdot)$.

\begin{de}\label{de:PSve}
For a real number $\varepsilon \in [0,1]$, we say that a symmetric key cryptosystem $\Sigma$ is  $\PSsm{\varepsilon}$--, $\PScs{\varepsilon}$--, or $\PScm{\varepsilon}$--secure if $\Sigma$ satisfies the following conditions: 
\begin{description}
\item[$\PSsm{\varepsilon}$:]
\hspace*{5mm} $\forall P_M\in\c{P}(\c{M})$,\\
\hspace*{10mm}  $\forall c \in \c{C}$, $d(P_{M|C}(\cdot|c),P_M(\cdot))\le \varepsilon$
\item[$\PScs{\varepsilon}$:]
\hspace*{5mm} $\forall P_M\in\c{P}(\c{M})$,\\
\hspace*{10mm}  $\forall m \in \c{M}$, $d(P_{C|M}(\cdot|m),P_C(\cdot))\le \varepsilon$
\item[$\PScm{\varepsilon}$:]
\hspace*{5mm} $\forall P_M\in\c{P}(\c{M})$,\\
\hspace*{10mm}   $d(P_{CM}(\cdot|\cdot),P_C(\cdot)P_M(\cdot))\le \varepsilon$ \QED
\end{description}
\end{de}
\medskip

As shown above,  $\PSsm{0}$, $\PScs{0}$ and $\PScm{0}$ are equivalent to \eqref{eq:PS}--\eqref{eq:PScm}, respectively, and they are all equivalent. 
In this paper, we are interested in relations among these 
security notions when $\varepsilon$ is positive and sufficiently small. The main results of this paper are summarized as follows:
\begin{itemize}
\item $\PSsm{\varepsilon}$ is the {\em strongest} among three security notions in Definition \ref{de:PSve}, which reflects the observation that $\PSsm{\varepsilon}$ is the most straightforward extension of  \eqref{eq:PS} in Definition \ref{de:PS}. 
\item Two security notions in Definition \ref{de:PSve} except for $\PSsm{\varepsilon}$ are  equivalent to each other, and they are essentially equivalent to the statistical versions of  indistinguishability and semantic security 
which will be introduced later. As a result, it is clarified that indistinguishability and semantic security are weaker security notions even if they are formulated in information theoretically secure setting. 
\end{itemize}

\section{Perfect Secrecy 
and  Indistinguishability}\label{sec:ind}
We reformulate the security notion of {\em indistinguishability} denoted by $\IND{\varepsilon}$ which is suitable for information theoretically secure setting. Then, we discuss the relation between $\IND{\varepsilon}$ and three notions of perfect secrecy presented in Definition \ref{de:PSve}. 

It is easy to see that \eqref{eq:PSm} is also represented as 
$\forall m_0, \forall m_1 \in \c{M}$, $\forall c \in \c{C}$, 
$P_{C|M}(c|m_0)=P_{C|M}(c|m_1)$\footnotemark
\footnotetext{According to \eqref{eq:fundamental-relation} and Theorem \ref{thm:fundamental}, 
perfect secrecy equivalent to 
$\forall m_0 \in \c{M}$, $\forall m_1 \in \c{M}$, $\forall c \in \c{C}$, $\pr{\enc(m_0,K)=c}=\pr{\enc(m_1,K)=c}$, which appears in \cite[Proposition 9.3--{\em 7}.]{DK-01}. 
}, which is equivalent to 
\begin{eqnarray}
\label{eq:d_PS_IND}
\forall m_0 ,\forall m_1 \in \c{M},~
d(P_{C|M}(\cdot|m_0),P_{C|M}(\cdot|m_1))=0. 
\end{eqnarray}
Note that \eqref{eq:d_PS_IND} implies that cryptograms corresponding to arbitrarily chosen messages $m_0$ and $m_1$ cannot be statistically distinguished. 

We now relax the condition given by \eqref{eq:d_PS_IND} using a real number $\varepsilon \in[0,1]$ such that 
\begin{eqnarray}
\label{eq:e_PS_IND}
\forall m_0 ,\forall m_1 \in \c{M},~ 
d(P_{C|M}(\cdot|m_0),P_{C|M}(\cdot|m_1))\le\varepsilon. 
\end{eqnarray}

According to the definition of variational distance, $d(P_X,P_Y) \le \varepsilon$ can be rewritten as
\begin{eqnarray}
\hspace{-5mm}\forall f: \c{A} \rightarrow \{0,1\},~
\bigl| \pr{f(X)=1} -  \pr{f(Y)=1}\bigr|\le \varepsilon
\end{eqnarray}
and hence, \eqref{eq:e_PS_IND} is equivalent to 
\begin{multline}\label{eq:f}
\forall m_0 \in \c{M},~ \forall m_1 \in \c{M},~ \forall f: \c{C} \rightarrow \{0,1\},\\
|\pr{f(C)=1\mid M=m_0} - \pr{f(C)=1|M=m_1}|\\
\le \varepsilon.
\end{multline}

Note that, \eqref{eq:f} is the definiton of {\em computational} 
indistinguishability if the function $f$ is restricted to the family of functions which can be computed in polynomial time \cite{GM-jcss84,G_FCv1-2001}. 
Hence, we introduce a security notion of {\em statistical} indistinguishability 
based on \eqref{eq:f} as follows. 
\begin{de}\label{de:S_IND} For a real number $\varepsilon\in[0,1]$, 
we say that a symmetric key cryptosystem $\Sigma$ is {\em statistically $\varepsilon$--indistinguishable} ($\IND{\varepsilon}$--secure, for short) if $\Sigma$ satisfies 
\eqref{eq:e_PS_IND} (and also \eqref{eq:f}). \QED
\end{de}
\begin{rem}
Statistical indistinguishability introduced by Dodis--Smith \cite{DS-tcc05} looks different from 
Definition \ref{de:S_IND}, but it is easy to show that they are essentially the same. \QED
\end{rem}

In the following, we clarify the relation among security notions in Definitions \ref{de:PSve} and \ref{de:S_IND}. 

\begin{thm}\label{thm:PSm/IND}
For an arbitrary $\varepsilon \in[0,1]$, 
a symmetric key cryptosystem $\Sigma$ is $\PScs{\varepsilon}$--secure iff  $\Sigma$ is $\IND{\varepsilon}$--secure. \QED
\end{thm}

\noindent
{\em Proof of Theorem \ref{thm:PSm/IND}:} Observe for every $m\in\c{M}$ that 
\begin{eqnarray}
\nonumber
&&d\left(P_{C|M}(\cdot|m),P_C(\cdot)\right)\\
\nonumber
&&=\frac{1}{2}\sum_{c \in \c{C}}\left|
P_{C|M}(c|m) - \sum_{m' \in \c{M}} P_{C|M}(c|m')P_M(m')
\right|\\
\nonumber
&&=\frac{1}{2}\sum_{c \in \c{C}}\left|
\sum_{m'\in \c{M}}P_M(m')\left\{P_{C|M}(c|m) - P_{C|M}(c|m')\right\}
\right|\\
\label{eq:INS/PS}
\end{eqnarray}

First, we show that  $\Sigma$ is $\PScs{\varepsilon}$--secure if $\Sigma$ is $\IND{\varepsilon}$--secure. In this case, we assume that $\forall m, \forall m'\in\c{M}$, 
$d(P_{C|M}(\cdot|m),P_{C|M}(\cdot|m'))\le \varepsilon$, and hence, from \eqref{eq:INS/PS} we have 
\begin{eqnarray}
\nonumber
&&d\left(P_{C|M}(\cdot|m),P_C(\cdot)\right)\\
\nonumber
&&\le
\frac{1}{2}\sum_{m'\in\c{M}}P_M(m')\sum_{c \in \c{C}}\left|
P_{C|M}(c|m) - P_{C|M}(c|m')\right|\\
\nonumber
&&= \sum_{m'\in \c{M}} P_M(m')d\bigl(P_{C|M}(\cdot|m), P_{C|M}(\cdot|m')\bigr)\\
\nonumber
&&\le \sum_{m' \in \c{M}} P_M(m') \, \varepsilon\\
&&= \varepsilon 
\end{eqnarray}
and hence $\Sigma$ is $\PScs{\varepsilon}$--secure.

We prove the converse. Suppose that $\Sigma$ is $\PScs{\varepsilon}$--secure. Substitute both $m=m_0$ and 
\begin{eqnarray}\label{eq:delta_f}
P_M(m')=\delta_{m_1}(m')\DEF\left\{
\begin{array}{cll}
1,&\mbox{if}~~m'=m_1\\
0,&\mbox{otherwise}
\end{array}
\right.
\end{eqnarray}
 into \eqref{eq:INS/PS}. Then, we obtain
\begin{eqnarray}
\nonumber
d(P_{C|M}(\cdot|m_0),P_C(\cdot)) &=&d(P_{C|M}(\cdot|m_0),P_{C|M}(\cdot|m_1))\\
&\le& \varepsilon. 
\end{eqnarray}
Hence, $\Sigma$ is $\IND{\varepsilon}$--secure
if it is $\PScs{\varepsilon}$--secure.  \QED

The next theorem implies an equivalence between $\IND{\varepsilon}$ and $\PScm{\varepsilon}$.

\begin{thm}\label{thm:PScm/IND} For an arbitrary $\varepsilon \in[0,1]$, 
a symmetric key cryptosystem $\Sigma$ is $\PScm{\varepsilon}$--secure if $\Sigma$ is $\IND{\varepsilon}$--secure. Conversely, if $\Sigma$ is $\PScm{\varepsilon}$--secure, 
it is $\IND{2\varepsilon}$--secure. 
\QED
\end{thm}

{\em Proof of Theorem \ref{thm:PScm/IND}:} This proof is essentially the same with Theorem \ref{thm:PSm/IND}. Observe that $d(P_{CM},P_CP_M)$ can be calculated as follows:
\begin{eqnarray}
\nonumber
\hspace{-6mm}d(&P_{CM}&(\cdot,\cdot),P_C(\cdot)P_M(\cdot)) \\
\nonumber
&=\frac{1}{2}& \sum_{c\in\c{C}}\sum_{m\in\c{M}}|P_{CM}(c,m)-P_C(c)P_M(m)|\\
\nonumber
&=\frac{1}{2}&\sum_{c\in\c{C}}\sum_{m\in\c{M}}P_M(m)\left|P_{C|M}(c|m)-P_C(c)\right|\\
\nonumber
&=\frac{1}{2}&\sum_{c\in\c{C}}\sum_{m\in\c{M}}P_M(m)\\
&\times&\hspace{-1mm}
\left|\sum_{m'\in\c{M}}P_M(m')
\left\{P_{C|M}(c|m)-P_{C|M}(c|m')\right\}\right|
\label{eq:d_CM}
\end{eqnarray}

We show that  $\Sigma$ is $\PScm{\varepsilon}$--secure if $\Sigma$ is $\IND{\varepsilon}$--secure. In this case, we have from \eqref{eq:d_CM} that 
\begin{eqnarray}
\nonumber
&&d(P_{CM}(\cdot,\cdot),P_C(\cdot)P_M(\cdot)) \\
\nonumber
&&\le \hspace{-2mm}
\sum_{m,m'\in\c{M}} \hspace{-2mm}
P_M(m)P_M(m')d(P_{C|M}(\cdot|m)
,P_{C|M}(\cdot|m'))\\
&&\le \varepsilon
\end{eqnarray}
if $\forall m, \forall m' \in \c{M}, d(P_{C|M}(\cdot|m) ,P_{C|M}(\cdot|m'))\le \varepsilon$. 
Hence, if $\Sigma$ is $\IND{\varepsilon}$--secure, it is also $\PScm{\varepsilon}$--secure.

Then, suppose that $\Sigma$ is $\PScm{\varepsilon}$--secure. Then, substituting 
\begin{eqnarray}\label{eq:P_m0m1}
P_M(m)=\left\{
\begin{array}{cll}
1/2,&\mbox{if}~~m=m_0~\mbox{or}~m=m_1\\
0,&\mbox{otherwise}
\end{array}
\right.
\end{eqnarray}
into \eqref{eq:d_CM}, it follows that 
\begin{eqnarray}
\hspace{-5mm}d(P_{CM}
,P_C
P_M
) &=&\frac{1}{2}d(P_{C|M}(\cdot|m_0),P_{C|M}(\cdot|m_1))
\le \varepsilon. 
\end{eqnarray}
Hence, $\Sigma$ is $\IND{2\varepsilon}$--secure
if it is $\PScm{\varepsilon}$--secure. 
\QED

We have proved that $\PScs{\varepsilon}$, $\PScm{\varepsilon}$, and $\IND{\varepsilon}$ are the same security notions. On the other hand, in section \ref{sec:gap}, we show an example that $\PSsm{\varepsilon}$ is stronger security notion than the others in the case of $\varepsilon >0$. 

\section{Perfect Secrecy and Semantic Security}
\label{sec:ss}

We consider the relation between perfect secrecy and semantic security in information theoretically secure setting. Here, $\IND{\varepsilon}$ also plays a crucial role. 
\begin{de}[Statistical semantic security, \cite{RW-it06}]\label{de:SS-p}
For every real number $\varepsilon\in[0,1]$ we say that a symmetric key cryptosystem $\Sigma=(P_K,\enc,\dec)$ is {\em statistically $\varepsilon$--semantic secure} ($\SS{\varepsilon}$--secure, for short) if, for an 
arbitrary distribution of a message $P_M \in \c{P}(\c{M})$ and for an arbitrary map $f: \c{C} \rightarrow \{0,1\}$, there exists a random variable $G_f$ that depends on $f$ but is independent of $M$, so that for every map $h: \c{M} \rightarrow \{0,1\}$, it holds that
\begin{eqnarray}
\label{eq:RW_def}
\bigr|\pr{f(C) = h(M)} - \pr{G_f = h(M)}\bigr| \le \varepsilon. 
\end{eqnarray}
\QED
\end{de}

Intuitively, Definition \ref{de:SS-p} implies that a cryptogram $C$ is almost useless to obtain any {\em one} bit information of a message $M$, since \eqref{eq:RW_def} implies that, in order to guess one bit information $h(M)$ of a message $M$, there is no difference between by using a cryptogram $C$ and a map $f$,  and by using $f$ only with a random coin.

\begin{rem}
In \cite{RW-it06}, $(t,\varepsilon)$--entropic security is defined if a symmetric key cryptosystem $\Sigma$ satisfies Definition \ref{de:SS-p} for {\em every message with min-entropy} $t$, and it is shown that the key length is reduced to $n-t+\omega(\log n)$ bits 
for $(t,n^{-\omega(1)})$--entropic security\footnotemark
\footnotetext{$f=\omega(g) \Leftrightarrow \forall \epsilon>0, \exists n_0, \forall n \ge n_0$, $g(n) \le \epsilon f(n)$. }. Hence, Definition \ref{de:SS-p} coincides with $(0,\varepsilon)$--entropic security. Furthermore, it is pointed out in \cite{RW-it06} that $(0,0)$--entropic security is equivalent to PS in Definition \ref{de:PS}. \QED
\end{rem}

We are interested in the relation between 
PS introduced in Definition \ref{de:PSve}, and 
statistical semantic security $\SS{\varepsilon}$ 
when $\varepsilon > 0$. 
To see this, we show the following relation between $\IND{\varepsilon}$ and $\SS{\varepsilon}$. 

\begin{thm}\label{thm:IND/SS}
For arbitrary $\varepsilon \in [0,1]$, if a symmetric key cryptosystem $\Sigma$ is $\IND{\varepsilon}$--secure, then $\Sigma$ is also $\SS{\varepsilon}$--secure. Conversely, 
if $\Sigma$ is $\SS{\varepsilon}$--secure, then it is also $\IND{4\varepsilon}$--secure. \QED

\end{thm}
\medskip
\noindent
{\em Proof of Theorem \ref{thm:IND/SS}:} First, we prove that $\Sigma$ is $\SS{\varepsilon}$--secure if $\Sigma$ is $\IND{\varepsilon}$--secure. This proof 
 is essentially the same with the proof appeared in \cite{G_FCv1-2001} under computationally secure setting. Let $M^*$ be a random variable of a message which is independent of the legitimate message $M$. Then, assume that the random variable $G_f$ is generated by $P_{C|M}(c|m)$ and $M^*$, i.e., 
we define that $G_f \DEF f(C^*)$ where 
$P_{C^*}(c) \DEF \sum_{m_1}P_{C|M}(c|m_1)P_{M^*}(m_1)$ for $c\in\c{C}$ and $m \in \c{M}$. 

Let us define an indicator function $\mathbb{I}_{f,h}:\c{C} \times \c{M} \rightarrow \{0,1\}$ for maps $f$ and $h$ such that 
\begin{eqnarray}
\mathbb{I}_{f,h}(c,m) = 
\left\{
\begin{array}{cll}
1, & \mbox{if}~~f(c)=h(m)\\
0, & \mbox{otherwise}.\\
\end{array}
\right.
\end{eqnarray}
Then, the left hand side of \eqref{eq:RW_def} can be evaluated as 
\begin{eqnarray}
\nonumber
&&\left|
\pr{f(C)=h(M)} - \pr{G_f=h(M)}
\right|\\
\nonumber
&&=\left|
\pr{f(C)=h(M)} - \pr{f(C^*)=h(M)}
\right|\\
\nonumber
&&=\left|
\sum_{c,m_0}
\mathbb{I}_{f,h}(c,m_0)
\left\{
P_{CM}(c,m_0) - P_{C^*M}(c,m_0)
\right\}
\right|
\\
\nonumber
&&=\left|
\sum_{c,m_0}
\mathbb{I}_{f,h}(c,m_0)P_{M}(m_0)
\left\{
P_{C|M}(c|m_0) - P_{C^*}(c)
\right\}
\right|\\
\nonumber
&&=
\left|
\sum_{m_0,m_1}
P_{M}(m_0)P_{M^*}(m_1)
\right.
\\
&&
\nonumber
~~~~~~ 
\left.
\times \sum_{c}
\mathbb{I}_{f,h}(c,m_0)\{P_{C|M}(c|m_0) - P_{C|M}(c|m_1)\}
\right|
\\
\nonumber
&&=
\left|
\sum_{m_0,m_1}
P_{M}(m_0)P_{M^*}(m_1) \times \biggr\{
\pr{f_{h,m_0}(C)=1|M=m_0}
\right.
\\
&&
~~~~~~~~~~~~~
-\pr{f_{h,m_0}(C)=1|M=m_1}
\biggl\}\biggl|,
\label{eq:IND->SS}
\end{eqnarray}
where $f_{h,m_0}: \c{C} \rightarrow \{0,1\}$ is defined by $f_{h,m_0}(c)=1$ iff $\mathbb{I}_{f,h}(c,m)=1$. Then, 
due to the definition of $\IND{\varepsilon}$ given by \eqref{eq:d_PS_IND}, it is easy to see that \eqref{eq:IND->SS} can be bounded from above by $\sum_{m_0,m_1\in\c{M}}P_{M}(m_0)P_{M^*}(m_1)\cdot\varepsilon=\varepsilon$. 

Conversely, we show that $\Sigma$ is $\IND{4\varepsilon}$--secure 
if $\Sigma$ is $\SS{\varepsilon}$--secure. 
Assuming that a symmetric key cryptosystem $\Sigma$ is $\SS{\varepsilon}$--secure, 
there exist an arbitrary $f:\c{C} \rightarrow \{0,1\}$ and a random variable $G_f$ that depends on $f$ but is independent of $M$, and \eqref{eq:RW_def} holds for an arbitrary $h: \c{M} \rightarrow \{0,1\}$. 

Now, letting $h$ be a map that always outputs $1$ for every $m\in\c{M}$, it holds for arbitrary $f:\c{C} \rightarrow \{0,1\}$ that 
\begin{eqnarray}
\bigl|\pr{f(C) = 1} - \pr{G_f = 1}\bigr| \le \varepsilon
\end{eqnarray}
which is equivalent to
\begin{eqnarray}
\bigl|\pr{f(C) = 0} - \pr{G_f = 0}\bigr| \le \varepsilon.
\end{eqnarray}
Hence, for $\ell\in\{0,1\}$, it holds that 
\begin{eqnarray}
\pr{f(C)=\ell}\ge \pr{G_f=\ell}-\varepsilon
\end{eqnarray}
Multiplying both sides by $\pr{h(M)=\ell}\ge 0$, we have 
\begin{multline}
\pr{f(C)=\ell }\pr{h(M)=\ell }\\
\ge\big(\pr{G_f=\ell }-\varepsilon\big)\pr{h(M)=\ell },
\end{multline}
and hence, it follows that
\begin{eqnarray}
\nonumber
\sum_{\ell \in \{0,1\}}{\sf Pr}\{f&(C&)=\ell \}\pr{h(M)=\ell }\\
\nonumber
&\ge&
\sum_{\ell \in\{0,1\}}\big(\pr{G_f=\ell }-\varepsilon\big)\pr{h(M)=\ell }\\
&\ge&
\pr{G_f=h(M)}-\varepsilon. 
\end{eqnarray}
From \eqref{eq:RW_def} we obtain 
\begin{eqnarray}
\nonumber
{\sf Pr}\{f(C)&=&h(M)\}
-\sum_{\ell \in \{0,1\}}\pr{f(C)=\ell }\pr{h(M)=\ell }\\
\nonumber
&\le&\pr{f(C)=h(M)}-\pr{G_f=h(M)}+\varepsilon\\
&\le& 2\varepsilon. 
\end{eqnarray}
Similarly, by evaluating the upper bound of $\pr{f(C)=\ell}$, $\ell \in \{0,1\}$, we have 
\begin{eqnarray}
\nonumber
&&\biggr|
\pr{f(C)=h(M)}\\
\label{eq:before_lemma}
&&-\sum_{\ell \in \{0,1\}}\pr{f(C)=\ell}\pr{h(M)=\ell}\biggr|\le 2\varepsilon
\end{eqnarray}
Applying Lemma \ref{lem:binary} in Appendix \ref{app:proof} to this inequality\footnotemark\footnotetext{Let $X$ and $Y$ in Lemma \ref{lem:binary}
 be $f(C)$ and  $h(M)$, respectively. }, it holds that
\begin{multline}
\Bigr|
\pr{f(C)=h(M)=1}\\
\label{eq:SS->IND1_2}
-\pr{f(C)=1}\pr{h(M)=1}\Bigr|\le \varepsilon.
\end{multline}
Since $P_M\in\c{P}(\c{M})$ is arbitrary, we set $P_M$ in the same way as  \eqref{eq:P_m0m1} for arbitrarily fixed $m_0,m_1\in\c{M}$, 
and let $h(m) = \delta_{m_0}(m)$ which is defined by \eqref{eq:delta_f}. 
Then, \eqref{eq:SS->IND1_2} becomes
\begin{eqnarray}
\nonumber
&&\pr{M=m_0}\biggr|
\nonumber
\pr{f(C)=1\mid M=m_0}\\
\nonumber
&&-\sum_{\ell\in\{0,1\}}\pr{f(C)=1\mid M=m_\ell}\pr{M=m_\ell}\biggr|\\
\nonumber
&&=\frac{1}{4}\biggr|
\pr{f(C)=1\mid M=m_0}-\pr{f(C)=1\mid M=m_1}\biggr|\\
&&\le \varepsilon.
\end{eqnarray}
Therefore, $d(P_{C|M}(\cdot|m_0),P_{C|M}(\cdot|m_1))\le 4\varepsilon$ is established for 
every $m_0,m_1\in\c{M}$. 
\QED

\section{A Gap between Perfect Secrecy and Indistinguishability, Semantic Security}\label{sec:gap}

We show an exmaple of a symmetric key cryptosystem $\Sigma$ that is 
$\IND{\varepsilon}$--secure (and hence, it is also $\PScs{\varepsilon}$-- and $\PScm{\varepsilon}$--secure) with {\em arbitrarily small} $\varepsilon >0$, while it is $\PSsm{\varepsilon'}$--secure with $\varepsilon' \ge 1/2$. This fact means that $\PSsm{\varepsilon}$ is stronger than the other security notions. We note that $\PSsm{\varepsilon}$ is a straightforward extension of Shannon's perfect secrecy given by \eqref{eq:PS} in Definition \ref{de:PS}. 

\begin{ex}
For an arbitrary  even integer $n$, define $\c{C}=\{c_1,c_2,\ldots,c_n\}$ and $\c{M}=\{m_1,m_2,\ldots,m_n\}$.  Then, consider the following $n \times n$ probability transition matrix corresponding to $P_{C|M}$ such that 
\begin{multline}
\label{eq:counter_ex}
\bb{P}_{C|M} \\
=\left[
\begin{array}{ccccc}
n^{-1} + \delta & n^{-1} - \delta & \cdots & n^{-1} + \delta & n^{-1} - \delta \\
n^{-1} - \delta & n^{-1} + \delta & \cdots & n^{-1} - \delta & n^{-1} + \delta \\
n^{-1}  & n^{-1}  & \cdots & n^{-1}& n^{-1}  \\
\vdots &\vdots &\ddots&\vdots &\vdots\\
n^{-1}  & n^{-1}  & \cdots & n^{-1}& n^{-1}  \\
\end{array}
\right]
\end{multline}
where $\delta = \varepsilon/2 \in (0,n^{-1}]$, and the $(i,j)$ element of $\bb{P}_{C|M}$ is equal to $P_{C|M}(c_i|m_j)$. From Theorem \ref{thm:fundamental}, 
note that there exists a symmetric key cryptosystem $\Sigma_{\rm ex}$ corresponding to \eqref{eq:counter_ex} since it is doubly stochastic.  

It is easy to check that $d(P_{C|M}(\cdot|m_i),P_{C|M}(\cdot|m_j))$ is equal to $0$ or  $2\delta$ ($=\varepsilon$) for each $m_i,m_j \in \c{M}$. Hence, $\bb{P}_{C|M}$ realizes a $\IND{\varepsilon}$--secure symmetric key cryptosystem (and hence, it is also $\PScs{\varepsilon}$--, and $\PScm{\varepsilon}$--secure). 

On the other hand, 
for uniformly distributed messages, i.e., $P_M(m_i) = 1/n$, $\forall m_i \in \c{M}$, it is easy to see that the the transition probability matrix $\bb{P}_{M|C}$ corresponding to a family of posteriori conditional probability distributions 
$\{P_{M|C}(m|c)\}_{c\in \c{C},m \in \c{M}}$ corresponds to the transposed matrix of $\bb{P}_{C|M}$. Hence, in this case 
\begin{eqnarray}
d(P_{M|C}(\cdot|c),P_M(\cdot))
=\left\{
\begin{array}{cll}
n\delta/2, &\mbox{if}~c=c_1 ~\mbox{or}~c_2\\
0, & \mbox{otherwise} 
\end{array}
\right. 
\end{eqnarray}
which implies that $\Sigma_{\rm ex}$ is $\PSsm{\varepsilon'}$--secure with\footnotemark
\footnotetext{Note that $d(P_{M|C}(\cdot|c),P_C(\cdot)) \le \varepsilon$ holds for {\em every} $P_M$ to ensure  $\PSsm{\varepsilon}$--secure cryptosystems.} $\varepsilon' \ge n \delta/2$. 
In particular, 
$\varepsilon' \ge 1/2$ for every $n$ if $\varepsilon = 2/n$ $(=2\delta)$ which can be arbitrarily small for sufficiently large $n$. \QED
\end{ex}
\medskip

In this example, the symmetric key cryptosystem $\Sigma_{\rm ex}$ given by \eqref{eq:counter_ex} violates $d(P_{M|C}(\cdot|c),P_M(\cdot)) \le \varepsilon$ with the negligibly small probability $\pr{C=c_1 \vee C=c_2}=2/n$ if $P_M$ is uniform and $n$ is sufficiently large, although it is required by $\PSsm{\varepsilon}$--security to satisfy $d(P_{M|C}(\cdot|c),P_M(\cdot)) \le \varepsilon$ for {\em every} $c \in\c{C}$. On the other hand, $\Sigma_{\rm ex}$ is still considered to be secure under the other security notions since they focus on the probability distribution of $C$ and the probability that such insecure cryptograms are output is negligible.

\medskip 

\section*{Acknowledgement}The authors would like to thank Prof.\ Hideki Imai in Chuo University, Prof.\ Ryutaroh Matsumoto in Tokyo Institute of Technology, and Mr.\ Yusuke Sakai in University of Electro-Communications for their helpful comments. The work of the first author, M.\ Iwamoto is partially supported
by the MEXT Grant-in-Aid for Young Scientists (B) No.\ 20760236.

\begin{table}[tb]
\caption{$P_{XY}$ and its marginals}
\begin{center}
\begin{tabular}{c|cc|c}
\hline
\hline
$x \backslash y$ &0&1& $P_X(x)$\\
\hline
$0$ &$a$&$b$& $a+b$\\
$1$ &$c$&$d$& $c+d$\\
\hline
$P_Y(y)$ &$a+c$&$b+d$& $1$\\
\hline
\hline
\end{tabular}
\end{center}
\label{table:A=2}
\end{table}%


\appendix

\subsection{Proof of Theorem \ref{thm:fundamental}}
\label{app:fundamental-relation}

Observe that a random variable $C$ of a cryptogram 
is obtained by $C=\enc(M,K)$, where $M$ and $K$ are independent random variables of a message and a key, respectively, and $\enc: \c{M} \times \c{K} \rightarrow \c{C}$ is a deterministic map of encryption. Hence, the joint probability distribution $P_{CM}(c,m)$ of a cryptogram and a message can be represented as 
\begin{eqnarray}
\nonumber
P_{CM}(c,m)&=&\pr{C=c,M=m}\\
\nonumber
&=&\pr{\enc(M,K)=c,M=m}\\
\nonumber
&=&\sum_{k:\enc(m,k)=c}P_{MK}(m,k)\\
\nonumber
&\stackrel{(*)}{=}&P_{M}(m)\sum_{k:\enc(m,k)=c}P_{K}(k)\\
&=&P_{M}(m)\pr{\enc(m,K)=c},
\end{eqnarray}
where the marked equality holds since $M$ and $K$ are independent. 
Hence, we have \eqref{eq:fundamental-relation}. 

In what follows, we consider the case of $|\c{M}|=|\c{C}|$. In this case, if $k \in \c{K}$ is fixed, there exists a bijection $\pi_k: \c{M}\rightarrow \c{C}$ since every cryptogram $c\in \c{C}$ can be uniquely decrypted by $k \in \c{K}$. Hence, for each $k \in \c{K}$, let $\Pi_k \in \{0,1\}^{n \times n}$ be a permutation matrix which corresponds to the bijection $\pi_k$. Then, it is easy to see that the probability transition matrix induced by $\enc$ and $K$ can be represented as 
\begin{eqnarray}\label{eq:decomposition}
\bb{P}_{C|M} = \sum_{k\in \c{K}}P_K(k) \Pi_k,
\end{eqnarray}
which is doubly stochastic. Conversely, due to Birkoff--von Neumann Theorem, there exists a pair of $P_K(k)$ and $\Pi_k$, $k \in \c{K}$, satisfying \eqref{eq:decomposition}  if $\bb{P}_{C|M}$ is doubly stochastic. \QED


\subsection{Lemma in Proof of Theorem \ref{thm:IND/SS}}
\label{app:proof}

In proof of Theorem \ref{thm:IND/SS}, we use the following lemma:

\begin{lem}\label{lem:binary}
For two binary random variables $X$ and $Y$ over a set $\{0,1\}$, 
and for $\varepsilon \in [0,1]$, the following two inequalities are equivalent:
\begin{eqnarray}
\label{eq:sum}
\left|\pr{X=Y} - \sum_{\ell \in \{0,1\}} \pr{X=\ell}\pr{Y=\ell}\right| &\le& \varepsilon
\\
\nonumber
\Bigr|\pr{X= Y=\ell} - \pr{X=\ell}\pr{Y=\ell}\Bigl|\, &\le& \frac{\varepsilon}{2},
\label{eq:resp}\\
\ell \in \{0,1\}&&
\end{eqnarray}
\QED
\end{lem}

We show that \eqref{eq:sum} $\Rightarrow$ \eqref{eq:resp} since \eqref{eq:resp} $\Rightarrow$ \eqref{eq:sum} is obvious. 
Letting $P_{XY}(x,y)$, $x, y\in \{0,1\}$ be a joint probability distribution of $X$ and $Y$ given by TABLE \ref{table:A=2}, 
\eqref{eq:sum} is equivalent to
\begin{eqnarray}\label{eq:abcd}
\bigr|a+d - (a+b)(a+c) - (c+d)(b+d)\bigr|\le \varepsilon. 
\end{eqnarray}
Since it holds that $a+b+c+d=1$, \eqref{eq:abcd} becomes 
$|ad-bc| \le \varepsilon/2$. Furthermore, using $a+b+c+d=1$ again, we have 
\begin{eqnarray}
\hspace{-7mm}
\bigl| P_{XY}(0,0) - P_X(0)P_Y(0)\bigr|
= \bigl|a-(a+b)(a+c)\bigr| 
&\le& \frac{\varepsilon}{2} \\
\hspace{-7mm}
\bigl| P_{XY}(1,1) - P_X(1)P_Y(1)\bigr|=\bigl|d- (c+d)(b+d)\bigr| 
&\le& \frac{\varepsilon}{2}  
\end{eqnarray}
which implies \eqref{eq:sum}. \QED

\begin{rem}
Note that \eqref{eq:sum} $\Rightarrow$ \eqref{eq:resp} does not generally hold if $X$ and $Y$ are not binary random variables. 
\QED
\end{rem}
\end{document}